\documentclass[draft]{agujournal2019}
\usepackage{url}
\usepackage{lineno}
\usepackage[inline]{trackchanges}
\usepackage{soul}
\usepackage{amsmath} 
\usepackage{amssymb}
\usepackage{graphicx}
\usepackage{threeparttable}
\renewcommand{\linenumbers}{}

\draftfalse

\journalname{JGR: Solid Earth}

\makeatletter
\def\ps@headings{\def\@oddfoot{\centerline{\small --\the\c@page--}}%
  \let\@evenfoot\@oddfoot\def\@oddhead{}\let\@evenhead\@oddhead}
\makeatother

\begin{document}

\title{Is Earthquake Replenishment Sufficient for Unbiased b-Value Estimation?}

\authors{H. H. Prabowo\affil{1}, G. Petrillo\affil{2,*}}

\affiliation{1}{Asian School of the Environment, Nanyang Technological University, Singapore}
\affiliation{2}{CNRS, ENS de Lyon, LPENSL, UMR5672, 69342, Lyon cedex 07, France}

\correspondingauthor{Giuseppe Petrillo}{giuseppe.petrillo@ens-lyon.fr}

\begin{keypoints}
\item Earthquake replenishment substantially improves conventional b-value estimates affected by short-term catalog incompleteness
\item Robust positive-difference estimators accurately recover the b-value without reconstructing missing earthquakes
\item Replenishment and robust estimation provide complementary rather than equivalent solutions to catalog incompleteness
\end{keypoints}

\begin{abstract}
Short-term aftershock incompleteness biases Gutenberg--Richter $b$-value estimates because many small earthquakes remain undetected immediately after large events. Two strategies address this problem. Earthquake replenishment reconstructs missing events to restore a more complete catalog, whereas robust positive-difference estimators infer the $b$-value directly from incomplete observations. Whether these approaches provide equivalent information on the magnitude distribution remains open. We address this issue using synthetic aftershock sequences generated with the Epidemic-Type Aftershock Sequence (ETAS) model and degraded through a time- and magnitude-dependent incompleteness model. Incomplete catalogs are replenished using AutoBETA, and $b$ is estimated using the standard maximum-likelihood, $b$-more-positive, and $b$-more-incomplete estimators. Replenishment reconstructs the overall structure of the missing-event population and substantially improves conventional maximum-likelihood estimates, although small residual biases remain. By contrast, both positive-difference estimators recover the imposed $b$-value directly from incomplete synthetic catalogs. Applications to the Ridgecrest, Landers, and Hector Mine sequences show that $b$-more-incomplete is less sensitive to replenishment than $b$-more-positive. Ten independent replenishment seeds indicate that seed-to-seed variability is small relative to differences among estimators. Replenishment and robust $b$-value estimation therefore address different aspects of catalog incompleteness: replenishment provides a more complete aftershock representation, whereas robust estimators are preferable when accurate $b$-value estimation is the primary objective.
\end{abstract}

\section*{Plain Language Summary}
Immediately after a large earthquake, many small aftershocks are not detected because their seismic signals overlap with those of larger events. This temporary loss of earthquakes can bias estimates of the Gutenberg--Richter $b$-value, a key parameter describing the relative occurrence of small and large earthquakes. One possible solution is to reconstruct the missing earthquakes and generate a replenished catalog. Another is to estimate the $b$-value using statistical methods specifically designed to remain accurate even when the catalog is incomplete. In this study, we compare these two approaches using realistic computer simulations and three well-known earthquake sequences from Southern California. We show that replenishment successfully restores much of the missing seismicity and significantly improves conventional estimates of the $b$-value. However, reconstructing a more complete catalog does not necessarily guarantee a completely unbiased estimate of the earthquake size distribution. Robust positive-difference estimators provide accurate $b$-values directly from the incomplete catalog, without requiring reconstruction of missing earthquakes. These results highlight that the two approaches are complementary and should be selected according to the scientific objective of the analysis.

\section{Introduction}
\label{sec:introduction}

The Gutenberg-Richter law provides one of the most robust empirical descriptions of earthquake occurrence, relating the frequency of earthquakes to their magnitude through
\begin{linenomath*}
\begin{equation}
N(M \geq m) = N_0 10^{-b(m-m_0)},
\label{eq:gr_intro}
\end{equation}
\end{linenomath*}
where $N(M \geq m)$ is the number of earthquakes with magnitude greater than or equal to $m$, $m_0$ is a lower reference magnitude, and $b$ controls the relative proportion of small and large earthquakes \cite{GR44}. Beyond its statistical role in characterizing the magnitude-frequency distribution, the $b$-value has been widely investigated as a potential indicator of the mechanical state of fault systems. Laboratory experiments, theoretical arguments, and observations from natural seismicity suggest that the $b$-value may depend on differential stress, material heterogeneity, and tectonic regime \cite{Sch68,Wys73,Sch15,godano2022dependence}. Spatial or temporal variations in $b$ have therefore been interpreted in terms of evolving stress conditions and have been explored in studies of earthquake preparation, aftershock evolution, and seismic hazard \cite{TWM14,tormann2015randomness,GW19,godano2024b,godano2024testing,godano2025bvalue,lippiello2025toward}. However, the physical interpretation of such variations requires that methodological and observational biases be carefully controlled \cite{MSST20}.

A principal source of bias is earthquake catalog incompleteness. Small earthquakes may be absent because of insufficient network coverage, unfavorable noise conditions, or limitations in phase detection and event association. These effects produce a spatially and temporally variable magnitude of completeness \cite{SW08,MWWCW11,MW12}. A particularly severe form of incompleteness occurs immediately after large earthquakes, when the waveforms of small aftershocks overlap with the coda of previous events. This phenomenon, commonly referred to as short-term aftershock incompleteness, causes the detection threshold to increase abruptly after a large earthquake and then decrease as the seismic rate decays \cite{Kag04,HKJ06,PVIH07,Hai16,Hai16a}. Because undetected events are concentrated primarily at low magnitudes, short-term incompleteness modifies the observed magnitude-frequency distribution and generally leads to an underestimation of the $b$-value \cite{dAGL18,PL20,LP24}.

The conventional strategy for limiting this bias is to estimate a completeness magnitude $M_c$ and apply the maximum-likelihood estimator only to earthquakes above a threshold $M_{\mathrm{th}} \geq M_c$ \cite{Aki65,OK93,GPL24}. This approach is effective when $M_c$ is approximately constant and can be estimated reliably. Its application becomes more problematic during the early stages of aftershock sequences, because $M_c$ can vary rapidly with time and depend on the entire recent seismic history. A conservative choice of $M_{\mathrm{th}}$ may reduce the completeness bias, but it also discards a large fraction of the catalog. The resulting reduction in sample size increases statistical uncertainty and limits the temporal and spatial resolution at which the $b$-value can be investigated \cite{SB82,GPL24}. Thus, estimating the $b$-value in incomplete catalogs requires balancing two competing objectives: reducing systematic bias while retaining enough earthquakes to obtain a statistically precise estimate.

Complementary approaches have recently sought to estimate temporal
$b$-value variations while explicitly accounting for statistical
uncertainty and changing detectability. \citeA{iwata2024adaptive} introduced
a state-space formulation estimated by particle filtering, which adapts
the temporal resolution of $b$-value estimates to the observed seismicity.
More recently, \citeA{laporte2025b} proposed the \emph{b}-Bayesian
framework, which jointly infers temporal changes in the $b$-value and
catalog detectability from non-truncated magnitude data. These approaches
address the inference of temporal $b$-value variations without explicitly
reconstructing the missing-event population. They are therefore
complementary to the replenishment strategy examined here.

Recent developments have addressed this problem through two conceptually distinct strategies. The first seeks to construct estimators that remain robust without explicitly reconstructing undetected earthquakes. The $b$-positive estimator introduced by \citeA{Van21} uses the distribution of positive magnitude differences between consecutive earthquakes. Its robustness follows from the conditional nature of the detection problem: once an earthquake of magnitude $m_i$ has been detected, a later earthquake with magnitude larger than $m_i$ is generally more likely to be recorded. \citeA{LP24} and \citeA{lippiello2026improving} generalized this idea by considering the first subsequent earthquake larger than a given event, thereby using a larger fraction of the observed catalog. The resulting $b$-more-positive estimator improves statistical efficiency, while the $b$-more-incomplete estimator further reduces residual detection biases by applying an artificial blind-time filter before evaluating the positive magnitude differences. These methods are designed to estimate the slope of the underlying magnitude distribution directly from incomplete observations, rather than to infer the individual earthquakes that were missed.

A second strategy aims to reconstruct the missing portion of the earthquake record itself. \citeA{zhuang2020detection,zheng2021data} developed a nonparametric replenishment framework for marked point processes with time-separable marks. The method is based on the biscale empirical probability integral transformation, which maps occurrence times and magnitudes onto a unit square. For a completely observed process with separable marks, the transformed events are approximately homogeneous; time and magnitude dependent missingness produces a depleted region in this transformed domain. The marginal empirical distributions are then iteratively corrected, and additional events are generated within the inferred missing region before being mapped back to the original time-magnitude domain. Automated implementations of this framework, including AutoBETA \cite{zhan2026autobeta}, make replenishment applicable to large earthquake catalogs and provide a practical tool for reconstructing the population of undetected aftershocks. Replenished catalogs can be valuable for studying seismicity rates, aftershock decay, triggering processes, and other quantities that depend on the number and occurrence times of earthquakes, in addition to the magnitude distribution itself.

Robust estimation and catalog replenishment therefore address short-term incompleteness from different directions. Positive-difference estimators seek to recover a specific statistical property, the Gutenberg-Richter slope, without assigning occurrence times or magnitudes to missing earthquakes. Replenishment instead seeks to recover a statistically plausible representation of the missing-event population, from which a range of seismological quantities can subsequently be estimated. Although both approaches rely on the assumption that the underlying magnitude distribution is separable from earthquake occurrence times, their objectives and sources of uncertainty are not identical. A replenished catalog may reproduce the main characteristics of the unobserved population without matching the original catalog event by event. Conversely, a robust $b$-value estimator may recover the magnitude-frequency slope accurately while providing no reconstruction of the missing seismicity. Whether these two strategies yield equivalent inferences for the $b$-value has not yet been systematically evaluated under controlled conditions.

In this study, we compare earthquake replenishment and robust positive-difference estimators specifically from the perspective of Gutenberg-Richter $b$-value estimation. We first generate complete synthetic aftershock sequences using the Epidemic-Type Aftershock Sequence model and impose realistic time and magnitude dependent short-term incompleteness following the ETAS incomplete formulation of \citeA{PL20}. Because the complete catalog and its theoretical $b$-value are known, the performance of each method can be evaluated directly. We then replenish the incomplete catalogs using AutoBETA and compare the standard maximum-likelihood estimate with the $b$-more-positive and $b$-more-incomplete estimates before and after replenishment. Finally, we apply the same analysis to the 1992 Landers, 1999 Hector Mine, and 2019 Ridgecrest sequences in Southern California. Our objective is not to assess replenishment as a general-purpose catalog reconstruction method, but to determine how reliably a replenished catalog preserves the magnitude-frequency distribution and to clarify when additional care is required if the principal scientific target is the $b$-value. 

\begin{figure}[htbp]
\centering
\includegraphics[width=\textwidth]{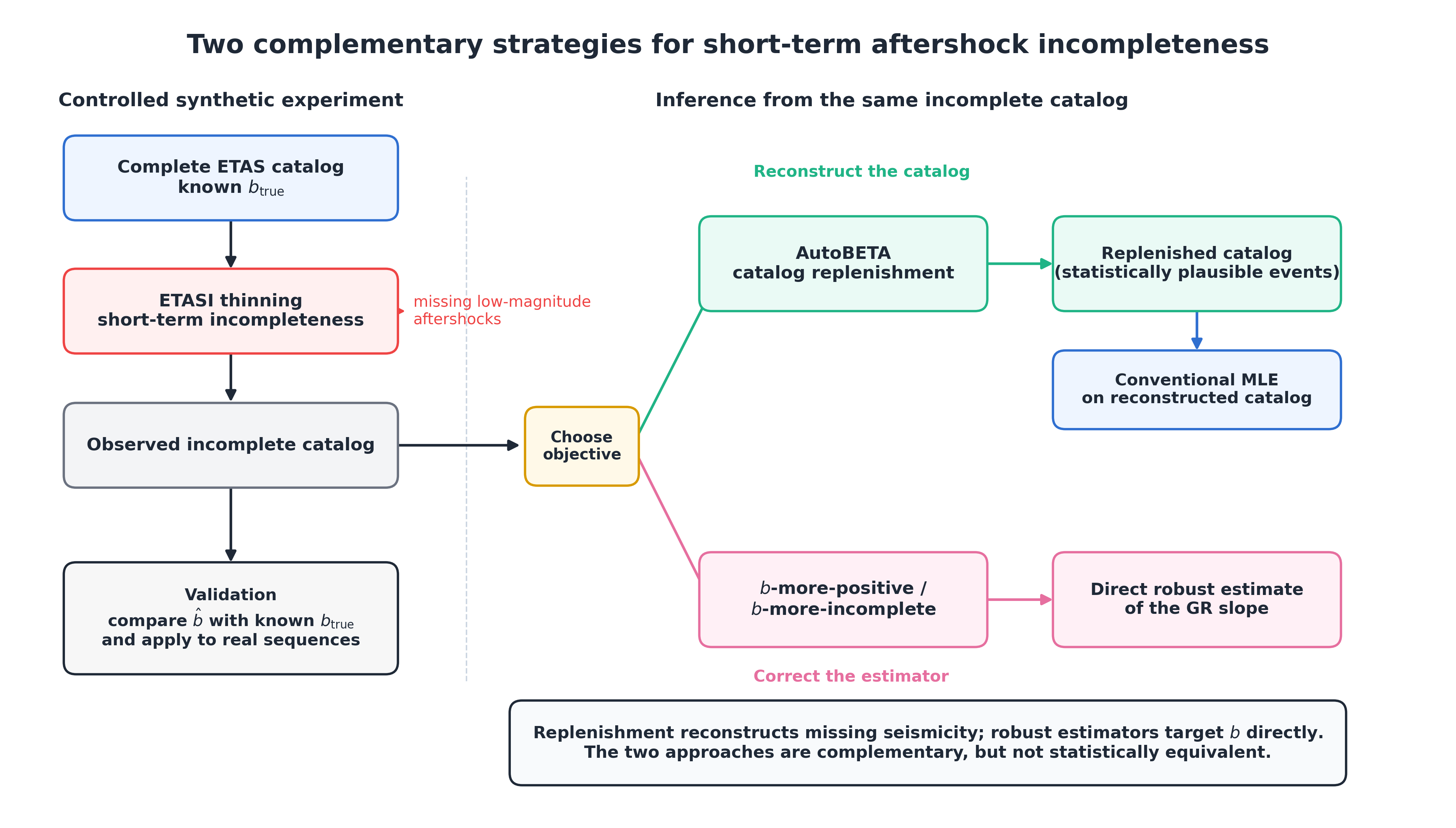}
\caption{
Conceptual workflow of the study. A complete synthetic ETAS catalog is first degraded using the ETASI short-term aftershock incompleteness model to produce an observed incomplete catalog. Two complementary strategies are then compared. The first reconstructs the missing seismicity through the AutoBETA replenishment algorithm before estimating the Gutenberg-Richter $b$-value using the conventional maximum-likelihood estimator. The second estimates the $b$-value directly from the incomplete catalog using the $b$-more-positive and $b$-more-incomplete estimators. Because the synthetic catalogs have a known theoretical $b$-value, the performance of both strategies can be evaluated quantitatively before applying the same workflow to real Southern California aftershock sequences.
}
\label{fig:workflow}
\end{figure}

Figure~\ref{fig:workflow} provides an overview of the workflow adopted throughout the paper and highlights the conceptual difference between reconstructing the missing seismicity and directly estimating the Gutenberg–Richter slope from incomplete observations.

\section{Data}
\label{sec:data}

\subsection{Synthetic Earthquake Catalogs}
\label{subsec:synthetic_data}

The synthetic data set consists of $N_{\mathrm{sim}}=500$ independent
aftershock sequences. Each realization was initiated from the same fixed
mainshock, with magnitude $M_{\mathrm{M}}=7.2$. The mainshock
was manually supplied as the initial event and was not sampled from a
background process. Occurrence times used in the analysis were measured
relative to the mainshock,
\begin{equation}
\widetilde{t}_i=t_i-t_{\mathrm{M}},
\end{equation}
so that the mainshock corresponds to $\widetilde{t}=0$.

The simulation end time was $T=1{,}199{,}352{,}960$~s, corresponding to
38 years from the origin of the simulation clock and approximately
37.99 years after the imposed mainshock.
Earthquake magnitudes were generated in the interval
$m_0\leq m\leq m_{\max}$, with $m_0=1.0$ and $m_{\max}=7.5$. All generated
events with $m\geq1.0$ were retained in the complete synthetic catalogs.
The imposed Gutenberg--Richter slope was
$b_{\mathrm{true}}=1$, corresponding to
$\beta=b_{\mathrm{true}}\ln(10)=2.3$.

The parameters controlling the simulated aftershock sequences are
summarized in Table~\ref{tab:synthetic_parameters}. Only triggered
aftershocks were generated. Parameters present in the simulation software
for multiple background events and foreshock generation were inactive and
therefore did not enter the catalogs analyzed here.

\begin{table}[htbp]
\caption{Parameters of the Synthetic Aftershock Catalogs}
\label{tab:synthetic_parameters}
\centering
\begin{tabular}{lll}
\hline
Component & Parameter & Value \\
\hline
Simulation & Number of realizations $N_{\mathrm{sim}}$ & 500 \\
Simulation & Final time $T$ & $1{,}199{,}352{,}960$ s \\
Magnitude & Lower bound $m_0$ & 1.0 \\
Magnitude & Upper bound $m_{\max}$ & 7.5 \\
Magnitude & $b_{\mathrm{true}}$ & 1.0 \\
Magnitude & $\beta$ & 2.3 \\
Productivity & $K_0$ & 0.1 \\
Productivity & $\alpha$ & 0.8 \\
Temporal kernel & $p$ & 1.2 \\
Temporal kernel & $c$ & 691.2 s \\
Spatial kernel & $d$ & 0.01 \\
Spatial kernel & $\gamma$ & 0.85 \\
Spatial kernel & $q$ & 1.506 \\
Branching process & Branching ratio & 0.4767 \\
Incompleteness & $w$ & 1.4 \\
Incompleteness & $\delta$ & 0.9 \\
Incompleteness & Transition width $\sigma$ & 0 \\
\hline
\end{tabular}
\end{table}

\subsection{Southern California Earthquake Catalogs}
\label{subsec:instrumental_data}

We analyzed the aftershock sequences of the 28 June 1992 Landers
($M_w7.3$), 16 October 1999 Hector Mine ($M_w7.1$), and 6 July 2019
Ridgecrest ($M_w7.1$) earthquakes. The instrumental catalogs were treated
as fixed observed inputs: replenishment modified neither the original
event parameters nor the selection windows, but added stochastic
realizations of events inferred to be missing. After applying the spatial, temporal, magnitude, and quality selections described below, the observed catalogs contained 16,749 earthquakes for
Landers, 16,636 for Hector Mine, and 14,308 for Ridgecrest. For each event,
the analysis used its occurrence time, reported magnitude, longitude,
and latitude. Magnitudes were retained down to the lowervanalysis threshold $m=1.0$. The catalogs were obtained from \citeA{HYS12}. No declustering was
performed because the objective was to analyze the complete temporal
evolution of each aftershock sequence. 

\section{Methods}
\label{sec:methods}

Our analysis was designed to compare two conceptually different approaches to short-term catalog incompleteness. The first reconstructs a statistically plausible population of undetected earthquakes through catalog replenishment. The second estimates the Gutenberg-Richter $b$-value directly from the observed catalog using positive magnitude differences that are weakly affected by detection losses. Synthetic catalogs provide a controlled experiment in which the complete catalog, the removed earthquakes, and the true $b$-value are known. The same estimators and replenishment procedure are subsequently applied to instrumental earthquake sequences.

\subsection{ETAS Simulations}
\label{subsec:etas}

We generated complete synthetic aftershock sequences using the Epidemic-Type Aftershock Sequence (ETAS) model \cite{Oga85,Oga98}. Each earthquake $i$ is characterized by its occurrence time $t_i$, magnitude $m_i$, and epicentral position $\mathbf{x}_i$. Because the simulations contained one externally imposed mainshock and no
additional background events, the conditional intensity used for
$t>t_{\mathrm{M}}$ was
\begin{linenomath*}
\begin{equation}
\lambda(m,\mathbf{x},t \mid \mathcal{H}_t)
=
\left[
\sum_{i:t_i<t}
K_0 10^{\alpha(m_i-m_0)}
g(t-t_i)
f(\mathbf{x}-\mathbf{x}_i \mid m_i)
\right]
p_{\mathrm{tr}}(m),
\label{eq:etas_intensity}
\end{equation}
\end{linenomath*}
where the imposed mainshock was included as the first triggering event.
Thus, the background term was $\mu(\mathbf{x})=0$ during the simulated
sequence.

Magnitudes were independently sampled from a Gutenberg--Richter
distribution truncated at $m_{\max}=7.5$,
\begin{linenomath*}
\begin{equation}
p_{\mathrm{tr}}(m)
=
\frac{
\beta\exp[-\beta(m-m_0)]
}{
1-\exp[-\beta(m_{\max}-m_0)]
},
\qquad
m_0\leq m\leq m_{\max},
\label{eq:gr_density}
\end{equation}
\end{linenomath*}
with
\begin{equation}
\beta=b_{\mathrm{true}}\ln(10)=2.3,
\qquad
b_{\mathrm{true}}=1.0
\end{equation}
The upper truncation changes the normalization of the magnitude density
but has a negligible numerical effect on the inferred slope because
$\exp[-\beta(m_{\max}-m_0)]\simeq3.2\times10^{-7}$.

The temporal triggering kernel follows the modified Omori law,
\begin{linenomath*}
\begin{equation}
g(t-t_i)
=
\frac{p-1}{c}
\left(
1+\frac{t-t_i}{c}
\right)^{-p},
\qquad t>t_i,
\label{eq:temporal_kernel}
\end{equation}
\end{linenomath*}
whereas the normalized spatial kernel is
\begin{linenomath*}
\begin{equation}
f(\mathbf{x}-\mathbf{x}_i \mid m_i)
=
\frac{q-1}
{\pi d\,10^{\gamma(m_i-m_0)}}
\left[
1+
\frac{\lVert\mathbf{x}-\mathbf{x}_i\rVert^2}
{d\,10^{\gamma(m_i-m_0)}}
\right]^{-q}.
\label{eq:spatial_kernel}
\end{equation}
\end{linenomath*}
The parameters $c$ and $p$ determine the temporal decay of the aftershock rate, while $d$, $\gamma$, and $q$ control the spatial extent and decay of triggering.

The simulation parameters are reported in
Table~\ref{tab:synthetic_parameters}. The productivity parameters were
$K_0=0.1$ and $\alpha=0.8$, the temporal parameters were $p=1.2$ and
$c=691.2$~s, and the spatial parameters were $d=0.01$, $\gamma=0.85$,
and $q=1.506$. These values produce a branching ratio of approximately
0.4767. The conditions $p>1$ and $q>1$ ensure normalization of the
temporal and spatial triggering kernels, respectively.

The resulting catalogs are referred to as the \emph{complete} catalogs. Their completeness is exact by construction for $m\geq m_0$, and they therefore provide the reference against which the incomplete and replenished catalogs can be evaluated.

\subsection{ETAS Incomplete (ETASI) Model}
\label{subsec:etasi}

Short-term aftershock incompleteness was introduced as a deterministic
post-generation thinning of each complete synthetic catalog. This
operation was separate from the forward ETAS simulation. For every
candidate event $i$, each earlier event $j$ defined a time-dependent
detection threshold
\begin{linenomath*}
\begin{equation}
q_j(\Delta t_{ij})
=
m_j
-
w\log_{10}
\left(
\frac{\Delta t_{ij}}{t_0}
\right)
+
\delta,
\qquad
\Delta t_{ij}=t_i-t_j>0,
\label{eq:individual_detection_threshold}
\end{equation}
\end{linenomath*}
where $t_0=1$~s makes the logarithm dimensionless. All occurrence times
were expressed in seconds, so this expression is numerically identical
to the implemented formula
$q_j=m_j-w\log_{10}(\Delta t_{ij})+\delta$.

The effective threshold for event $i$ was the largest threshold produced
by the preceding events,
\begin{linenomath*}
\begin{equation}
q_{\mathrm{c},i}
=
\max_{j:t_j<t_i}
q_j(t_i-t_j).
\label{eq:effective_detection_threshold}
\end{equation}
\end{linenomath*}
Because the implemented transition width was zero, thinning was a hard deterministic operation,
rather than a random draw from a smooth detection probability. The
detection indicator was therefore
\begin{linenomath*}
\begin{equation}
D_i
=
\begin{cases}
1, & m_i\geq q_{\mathrm{c},i},\\
0, & m_i<q_{\mathrm{c},i},
\end{cases}
\label{eq:detection_indicator}
\end{equation}
\end{linenomath*}
where $D_i=1$ denotes a retained event. Equivalently,
\begin{equation}
D_i
=
\prod_{j:t_j<t_i}
H\!\left[
m_i-q_j(t_i-t_j)
\right],
\end{equation}
where $H$ is the Heaviside step function. The thresholds were evaluated
from the chronology of the complete generated catalog before the
undetected events were removed.

The thinning parameters were
\begin{equation}
w=1.4,
\qquad
\delta=0.9,
\qquad
\sigma=0.
\end{equation}
Thus, the detection threshold decreases by 1.4 magnitude units for each
tenfold increase in elapsed time. At $\Delta t=1$~s, an event of magnitude
$m_j$ produces the threshold $q_j=m_j+0.9$. We refer to events with
$D_i=1$ as the observed or thinned synthetic catalog and to events with
$D_i=0$ as the known missing population.

\subsection{Earthquake Replenishment}
\label{subsec:replenishment}

Missing earthquakes were reconstructed using the Auto-Biscale Empirical Transformation Application (AutoBETA) developed by \citeA{zhan2026autobeta}. AutoBETA is an automated implementation of the nonparametric replenishment framework introduced by \citeA{zhuang2020detection} and further applied to earthquake sequences by \citeA{zheng2021data}. The procedure assumes that, in the underlying complete process, earthquake magnitudes are statistically separable from occurrence times. It does not require a prescribed Gutenberg--Richter distribution, an Omori law, or an ETAS conditional intensity.

Let the observed catalog contain $n$ earthquakes with ordered occurrence times $t_i$ and magnitudes $m_i$. AutoBETA initially transforms each event into the unit square according to
\begin{linenomath*}
\begin{equation}
u_i=\widehat{F}_t(t_i),
\qquad
v_i=\widehat{F}_m(m_i),
\label{eq:bepit_transform}
\end{equation}
\end{linenomath*}
where $\widehat{F}_t$ and $\widehat{F}_m$ are the empirical cumulative distribution functions of occurrence time and magnitude. This mapping is the biscale empirical probability integral transformation. If the complete process has separable time and magnitude distributions, the transformed events are expected to be approximately homogeneous over $[0,1]^2$. Missing earthquakes instead produce a region of anomalously low point density.

AutoBETA divides the transformed domain into a regular grid and identifies low-density cells using a percentile-based criterion. Neighboring deficient cells are joined to form an initial estimate of the missing region, denoted by $S^{(0)}$. The marginal distributions are then corrected for the events inferred to be missing, and the transformation in Equation~\ref{eq:bepit_transform} is recomputed. This produces an updated region $S^{(1)}$. The procedure is iterated,
\begin{linenomath*}
\begin{equation}
S^{(0)}
\rightarrow
S^{(1)}
\rightarrow
\cdots
\rightarrow
S^{(r)}
\rightarrow
S^{*},
\label{eq:autobeta_iteration}
\end{equation}
\end{linenomath*}
until either the displacement of the missing-region boundary falls below the convergence threshold or the maximum number of iterations is reached. The final region $S^{*}$ represents the part of the transformed domain inferred to be depleted by catalog incompleteness.

Under homogeneity in the complete transformed domain, the expected number of missing events is related to the area of $S^{*}$. AutoBETA accounts for uncertainty in this number by sampling the replenished population from a negative binomial distribution. Conditional on the inferred missing region, new points are generated within $S^{*}$ and subsequently mapped back to the original time--magnitude domain using the inverse corrected marginal distributions,
\begin{linenomath*}
\begin{equation}
t_{\mathrm{new}}
=
\widehat{F}_t^{-1}(u_{\mathrm{new}}),
\qquad
m_{\mathrm{new}}
=
\widehat{F}_m^{-1}(v_{\mathrm{new}}).
\label{eq:inverse_bepit}
\end{equation}
\end{linenomath*}
The resulting combination of observed and generated events is referred to as the \emph{replenished catalog}.

For all synthetic and instrumental catalogs, we used the fully automated direct-iteration mode. The AutoBETA settings were a grid of $70 \times 70$ cells (for the instrumental sequences, $85\times85$ for Landers, and $80\times80$ for Hector Mine), a density percentile of $10$, a minimum connected-region size of $200$ cells, a convergence threshold of $10^{-6}$, and a maximum of $100$ iterations. The same parameter-selection rule was applied to all catalogs to avoid tuning the replenishment specifically for the desired $b$-value. AutoBETA returned a valid depleted-region polygon for 484 of the 500
thinned synthetic catalogs. For 16 realizations (3.2\%), the BEPIT
procedure did not identify a valid polygon satisfying the prescribed
geometrical criteria, and no replenished catalog could therefore be
generated. These realizations were not replaced or assigned a
replenishment result. Complete- and thinned-catalog statistics were
evaluated using all 500 realizations, whereas statistics involving
replenished catalogs were evaluated using the 484 successful
realizations.

AutoBETA reconstructs a statistically plausible missing population rather than the identities of the earthquakes removed from a particular synthetic realization. Consequently, an exact event-by-event match between the removed and replenished populations is neither expected nor required. Our assessment therefore focuses on whether the replenished population restores the statistical magnitude distribution and, in particular, the Gutenberg--Richter slope.

\subsection{$b$-Value Estimators}
\label{subsec:b_estimators}

We evaluated the Gutenberg-Richter slope using three estimators: the conventional maximum-likelihood estimator, the $b$-more-positive estimator, and the $b$-more-incomplete estimator. Each estimator was applied independently to the complete, thinned, and replenished synthetic catalogs and to the original and replenished instrumental catalogs.

The synthetic ETAS catalogs contain continuous magnitudes and therefore
require no de-binning. Magnitudes in the instrumental catalogs are
reported with a finite resolution $\Delta m$. Before applying any of the
three estimators, including the conventional Aki estimator, each
instrumental magnitude was converted to a continuous value according to
\begin{linenomath*}
\begin{equation}
m_i^{*}
=
m_i+r_i\Delta m,
\qquad
r_i\sim U(0,1).
\label{eq:debinning}
\end{equation}
\end{linenomath*}
This randomization is equivalent, in expectation, to applying the usual
half-bin correction when the reported magnitude is treated as the lower
edge of its discretization interval. The same de-binned magnitudes were
used consistently by the maximum-likelihood, $b$-more-positive, and
$b$-more-incomplete estimators.

\subsubsection{Maximum-Likelihood Estimator}

For a catalog represented by continuous magnitudes and assumed complete
above a magnitude threshold $M_{\mathrm{th}}$, the conventional Aki
maximum-likelihood estimator is \cite{Aki65}
\begin{linenomath*}
\begin{equation}
\widehat{b}_{\mathrm{MLE}}(M_{\mathrm{th}})
=
\frac{1}
{\ln(10)
\left(
\overline{m}-M_{\mathrm{th}}
\right)},
\label{eq:b_mle}
\end{equation}
\end{linenomath*}
where $\overline{m}$ is the average magnitude of the earthquakes satisfying
$m_i\geq M_{\mathrm{th}}$. Synthetic ETAS magnitudes are continuous by
construction. Instrumental magnitudes were converted to continuous
values through the de-binning procedure described below before applying
Equation~\ref{eq:b_mle}.

For the synthetic catalogs, we evaluated Equation~\ref{eq:b_mle} at
\begin{linenomath*}
\begin{equation}
M_{\mathrm{th}}\in\{1.0,1.5,2.0\},
\label{eq:mle_thresholds}
\end{equation}
\end{linenomath*}
thereby quantifying the dependence of the conventional estimate on the imposed lower threshold. These fixed thresholds are not intended as independent estimates of the time-varying completeness magnitude; rather, they provide a controlled measure of how much residual incompleteness remains at different magnitude levels.

\subsubsection{$b$-More-Positive Estimator}

The $b$-more-positive estimator was recently introduced by \citeA{LP24} as an extension of the $b$-positive estimator of \citeA{Van21}. Rather than using only consecutive earthquakes, the method exploits a much larger fraction of the catalog by searching forward from each earthquake until a suitable magnitude pair is identified.

For every earthquake $i$, the algorithm scans the subsequent catalog and identifies the first earthquake $j$ satisfying the following conditions:

\begin{enumerate}
\item the magnitude difference exceeds a prescribed threshold,
\begin{equation}
m_j-m_i \geq \delta M_{\mathrm{th}},
\end{equation}

\item no intermediate earthquake has a magnitude larger than $m_i$,
\begin{equation}
m_k<m_i,
\qquad
i<k<j,
\end{equation}

\item the temporal separation satisfies
\begin{equation}
t_j-t_i \leq \tau_{\max},
\end{equation}

\item the hypocentral (or epicentral) distance satisfies
\begin{equation}
r_{ij}\leq R_{\max},
\end{equation}

\item the search is terminated after at most $l$ subsequent earthquakes.
\end{enumerate}

The corresponding positive magnitude difference is then defined as

\begin{linenomath*}
\begin{equation}
\delta m_i=m_j-m_i.
\end{equation}
\end{linenomath*}

The rationale behind this construction is that, once an earthquake of magnitude $m_i$ has been detected, a subsequent earthquake with a larger magnitude is considerably less affected by short-term detection incompleteness. Consequently, the distribution of positive magnitude differences approximately follows the same exponential law as the underlying Gutenberg--Richter distribution \cite{Van21,LP24}.

Assuming Gutenberg--Richter statistics, the $b$-value is estimated as

\begin{linenomath*}
\begin{equation}
\widehat{b}_{++}
=
\frac{1}
{\ln(10)
\left(
\overline{\delta m}
-
\delta M_{\mathrm{th}}
\right)},
\label{eq:b_more_positive}
\end{equation}
\end{linenomath*}

where $\overline{\delta m}$ denotes the average of all selected positive magnitude differences.

Following \citeA{LP24}, the maximum search length was fixed to

\[
l=100,
\]

while the temporal and spatial search windows were chosen as

\[
\tau_{\max}={86,400s},
\qquad
R_{\max}={60km}.
\]

The magnitude threshold

\[
\delta M_{\mathrm{th}}
=
{0.2}
\]

was selected following the stabilization criterion proposed by \citeA{LP24}, namely as the smallest threshold beyond which the estimated $b$-value remained approximately constant.

\subsubsection{$b$-More-Incomplete Estimator}

Although the $b$-more-positive estimator is substantially less sensitive to short-term incompleteness than the conventional estimator, a residual bias may arise when the transition from undetected to fully detected earthquakes is
gradual. In this case, some earthquakes below the effective completeness level may still be recorded, and the probability of detecting the subsequent larger
earthquake is no longer approximately equal to one. To mitigate this effect, \citeA{LP24} introduced the $b$-more-incomplete estimator.

The method first applies an artificial blind-time filter to the observed catalog. For every earthquake, all subsequent earthquakes occurring within a prescribed blind time $\tau_b$ and having a smaller magnitude are discarded. The filtered catalog therefore satisfies

\begin{linenomath*}
\begin{equation}
\mathcal{C}_{\tau_b}
=
\left\{
i:
\nexists\,j<i
\text{ such that }
0<t_i-t_j<\tau_b
\text{ and }
m_j>m_i
\right\}.
\end{equation}
\end{linenomath*}

The same pairing algorithm described above is then applied to the filtered catalog, using

\[
\delta M_{\mathrm{th}}=0.
\]

The resulting estimator becomes

\begin{linenomath*}
\begin{equation}
\widehat{b}_{+\mathrm{inc}}
=
\frac{1}
{\ln(10)\,
\overline{\delta m}_{\tau_b}},
\label{eq:b_more_incomplete}
\end{equation}
\end{linenomath*}

where $\overline{\delta m}_{\tau_b}$ is the average positive magnitude difference computed after the blind-time filtering.

The blind time was selected according to the stabilization criterion proposed by \citeA{LP24}. In this study we adopted

\[
\tau_b
=
{200s}.
\]

\subsection{Statistical Evaluation}
\label{subsec:statistical_evaluation}

The estimators were evaluated for the complete and ETASI-thinned catalogs
of all 500 synthetic realizations. Estimates involving AutoBETA-
replenished catalogs were available for the 484 realizations for which
the BEPIT procedure returned a valid depleted-region polygon. All
replenished-catalog means, standard deviations, and probability densities
reported below are therefore based on these 484 successful realizations. Because the true value is known, performance was quantified through the relative error
\begin{linenomath*}
\begin{equation}
\epsilon_t
=
\frac{\widehat{b}}{b_{\mathrm{true}}}-1.
\label{eq:relative_b_error}
\end{equation}
\end{linenomath*}
Thus, $\epsilon_t=0$ indicates exact recovery of the imposed Gutenberg--Richter slope, whereas negative and positive values indicate underestimation and overestimation, respectively. The distributions of $\epsilon_t$ over the independent ETAS realizations were used to compare systematic bias and statistical dispersion among the estimators.

For estimators applied to a fixed catalog, bootstrap standard errors were
computed from 100 nonparametric replicates. For the synthetic experiment,
the reported distributions of $\epsilon_t$ summarize the mean and
standard deviation across the $N_{\mathrm{sim}}=500$ independent ETAS
realizations.

For the instrumental catalogs, stochastic replenishment uncertainty was
evaluated separately using the $S=10$ independent AutoBETA draws described
in Section~\ref{subsec:stochastic_stability}. We distinguish this
between-draw variability from the bootstrap uncertainty conditional on a
particular catalog realization.

\section{Synthetic Experiments}
\label{sec:synthetic_results}

\subsection{Simulated Incomplete Catalogs}
\label{subsec:synthetic_incomplete}

\begin{figure}[htbp]
\centering
\includegraphics[width=0.96\textwidth]{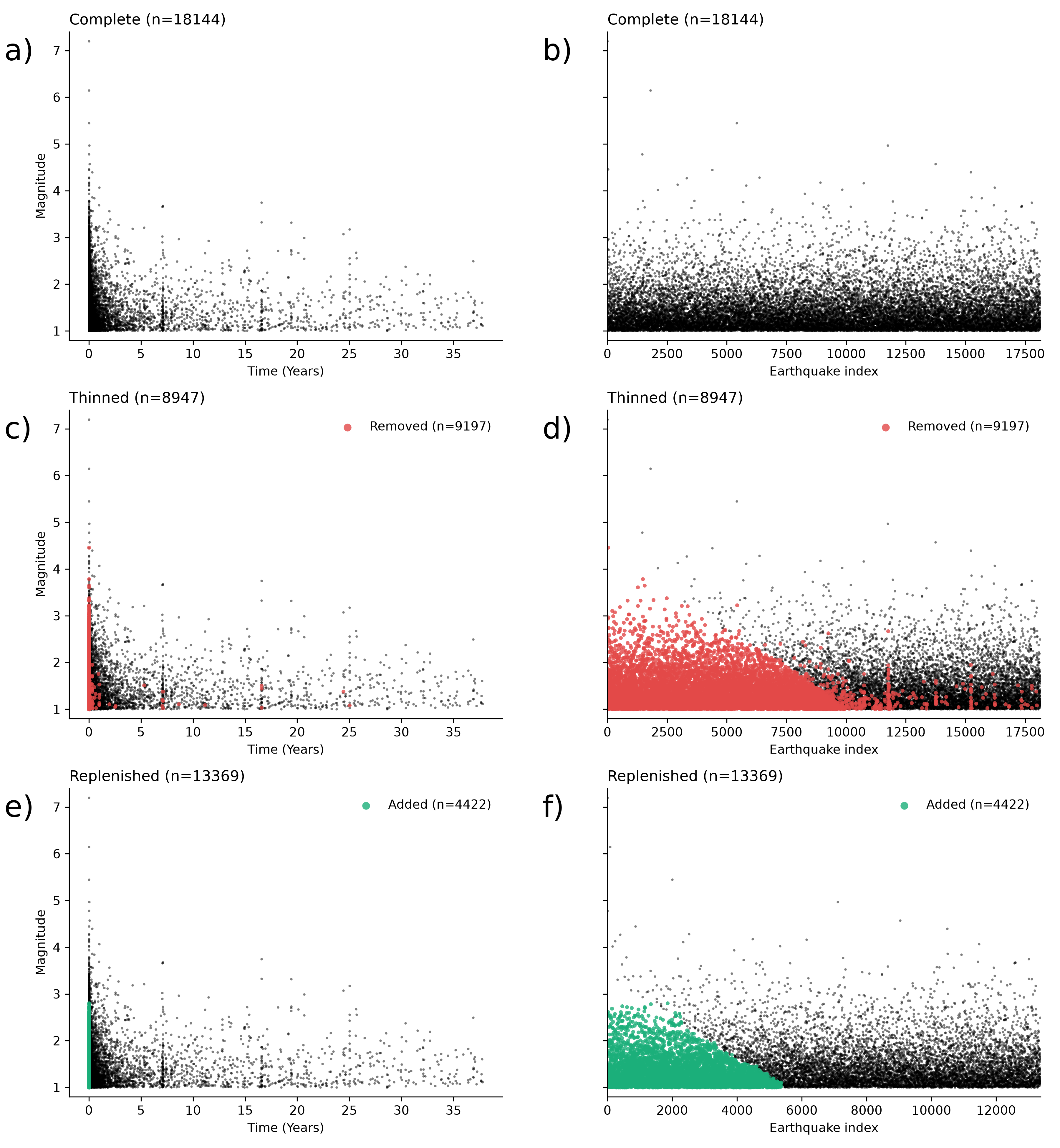}
\caption{Representative synthetic aftershock sequence generated by the ETAS model and degraded using the ETASI incompleteness model. Panels (a) and (b) show the complete catalog as a function of occurrence time and earthquake index, respectively. Panels (c) and (d) show the thinned catalog; earthquakes removed by the ETASI detection process are highlighted in red. Panels (e) and (f) show the AutoBETA-replenished catalog; newly generated earthquakes are highlighted in green. The complete, thinned, and replenished catalogs contain 18,144, 8,947, and 13,369 earthquakes, respectively.}
\label{fig:synthetic_catalogs}
\end{figure}

The controlled synthetic experiment was designed to separate the effects of short-term catalog incompleteness from the intrinsic statistical variability of the $b$-value estimators. For each ETAS realization, the initially complete catalog contains all earthquakes with $m\geq m_0=1$. The ETASI detection process was then applied to the same realization to remove earthquakes according to their magnitude and to the recent seismic history. The removed events are therefore known exactly and can be compared with the events subsequently generated by AutoBETA. \\
Figure~\ref{fig:synthetic_catalogs} illustrates one representative
realization. The complete catalog contains 18,144 earthquakes above the
lower magnitude threshold. Short-term incompleteness removes 9,197
earthquakes, leaving 8,947 events in the thinned catalog. The corresponding
missing fraction is
\begin{equation}
f_{\mathrm{miss}}
=
\frac{9{,}197}{18{,}144}
=
0.507,
\end{equation}
consistent with the approximately 50\% incompleteness level targeted when
selecting the thinning parameters. The removed population is concentrated
primarily at low magnitudes and during the early, high-rate portion of
the sequence.

AutoBETA adds 4,422 earthquakes, producing a replenished catalog
containing 13,369 events. We define the event recovery fraction as the
number of generated events divided by the known number of removed events,
\begin{equation}
f_{\mathrm{rec}}
=
\frac{N_{\mathrm{added}}}{N_{\mathrm{removed}}}
=
\frac{4{,}422}{9{,}197}
=
0.481.
\end{equation}
Thus, AutoBETA reconstructs approximately 48\% of the missing population
in this realization. The numerical similarity between
$f_{\mathrm{miss}}$ and $f_{\mathrm{rec}}$ is coincidental:
$f_{\mathrm{miss}}$ is controlled by the prescribed thinning model,
whereas $f_{\mathrm{rec}}$ is an output of the replenishment procedure.
The generated earthquakes nevertheless occupy the same general
time-magnitude region as the known missing population.

The comparison in Figure~\ref{fig:synthetic_catalogs} should not be interpreted as an event-by-event validation of replenishment. AutoBETA estimates a statistically plausible missing population using only the observed catalog and does not attempt to identify the specific earthquakes removed from the original ETAS realization. The relevant question for the present study is therefore whether the generated population sufficiently restores the magnitude-frequency distribution to permit an unbiased estimate of its slope.

\subsection{AutoBETA Reconstruction of the Missing-Event Region}
\label{subsec:synthetic_replenishment}

\begin{figure}[htbp]
\centering
\includegraphics[width=0.98\textwidth]{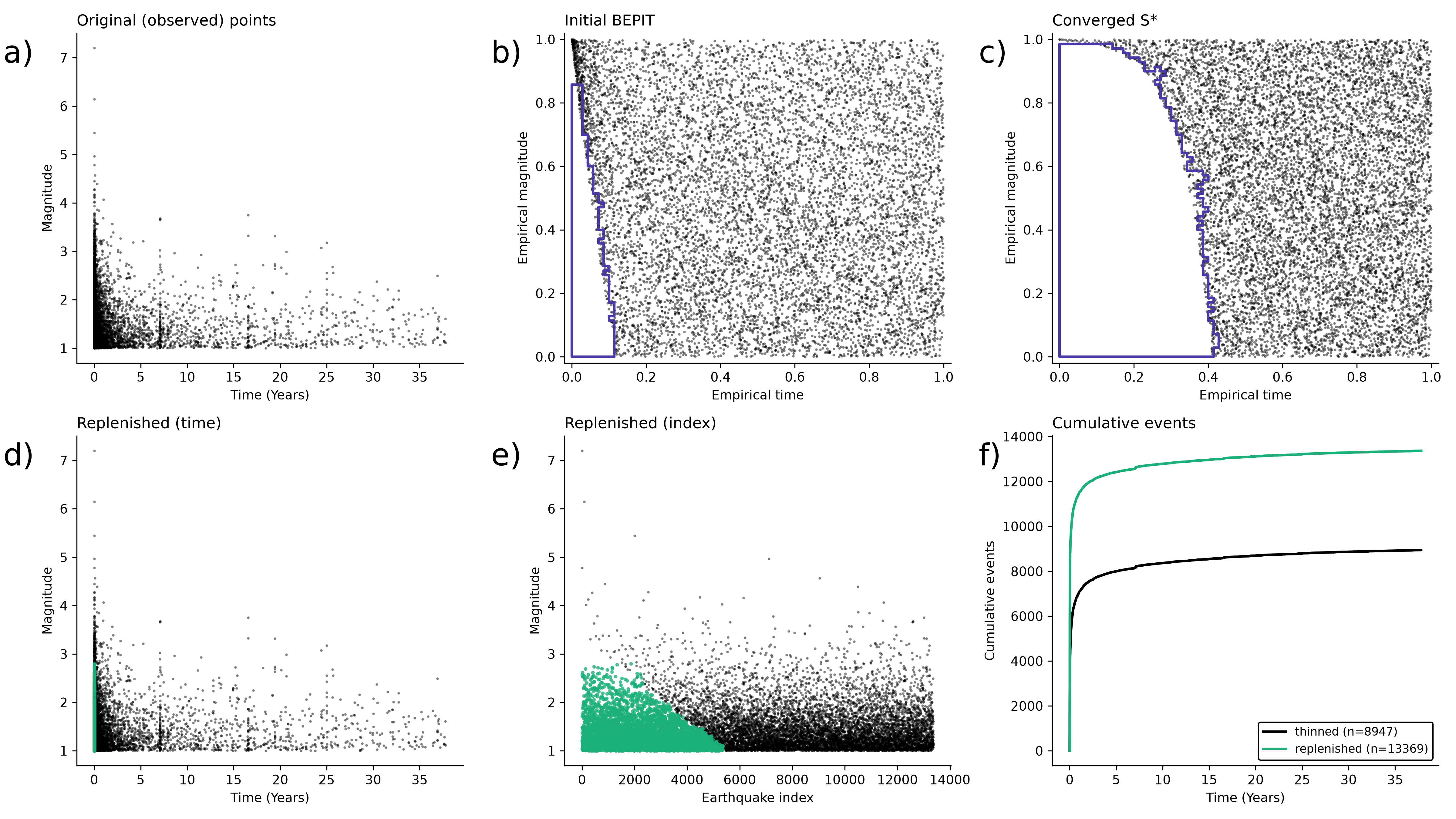}
\caption{Illustration of the AutoBETA replenishment procedure for a representative synthetic catalog. (a) Original thinned catalog in the time--magnitude domain. (b) Initial biscale empirical probability integral transformation, with the first estimate of the depleted region shown by the blue boundary. (c) Converged missing region $S^{*}$ after iterative correction of the empirical marginal distributions. (d) Replenished catalog as a function of occurrence time. (e) Replenished catalog as a function of earthquake index, with generated events shown in green. (f) Cumulative number of earthquakes in the thinned and replenished catalogs.}
\label{fig:autobeta_workflow}
\end{figure}

Figure~\ref{fig:autobeta_workflow} illustrates the main steps of the AutoBETA procedure for the same synthetic realization. The original catalog displays the characteristic rapid decay of seismicity following the mainshock together with a large deficit of low-magnitude earthquakes during the earliest part of the sequence. After the initial biscale empirical probability integral transformation, the observed points are mapped to the unit square. Catalog incompleteness appears as a depleted region near the lower-left part of this domain, where early occurrence times coincide with low empirical magnitudes.

The boundary inferred from the initial transformation provides a first estimate of the missing region. The empirical time and magnitude distributions are then iteratively corrected, and the transformed catalog is recalculated. During this procedure, the boundary changes from the initial estimate to the converged region $S^{*}$. The latter extends over a substantially larger fraction of the transformed domain and accounts for the fact that the initially observed empirical distributions are themselves distorted by missing earthquakes.

Once convergence is reached, additional transformed points are generated inside $S^{*}$ and mapped back to the original time--magnitude coordinates. In the time domain, the replenished events are concentrated mainly during the earliest part of the sequence. In the event-index representation, they reconstruct the declining envelope of the missing low-magnitude population. The cumulative event curve further shows that the principal effect of replenishment occurs immediately after the mainshock, although smaller corrections continue throughout the considered time interval.

The reconstructed population reproduces the broad geometry of the missing-event region but remains smaller than the population removed by ETASI. This difference is not necessarily indicative of a failure of the replenishment procedure, because the method estimates the missing population from the internal statistical structure of the incomplete catalog rather than from knowledge of the complete realization. It does, however, provide a particularly useful controlled setting in which to determine whether partial restoration of the missing population is sufficient for recovering the correct Gutenberg--Richter slope.

\subsection{Comparison of $b$-Value Estimators}
\label{subsec:synthetic_b_comparison}

We next compare the behavior of the conventional maximum-likelihood estimator, the $b$-more-positive estimator, and the $b$-more-incomplete estimator. For every synthetic realization and catalog state, the relative error was computed as

\begin{linenomath*}
\begin{equation}
\epsilon_t
=
\frac{\widehat{b}}{b_{\mathrm{true}}}-1.
\end{equation}
\end{linenomath*}

\begin{figure}[htbp]
\centering
\includegraphics[width=0.82\textwidth]{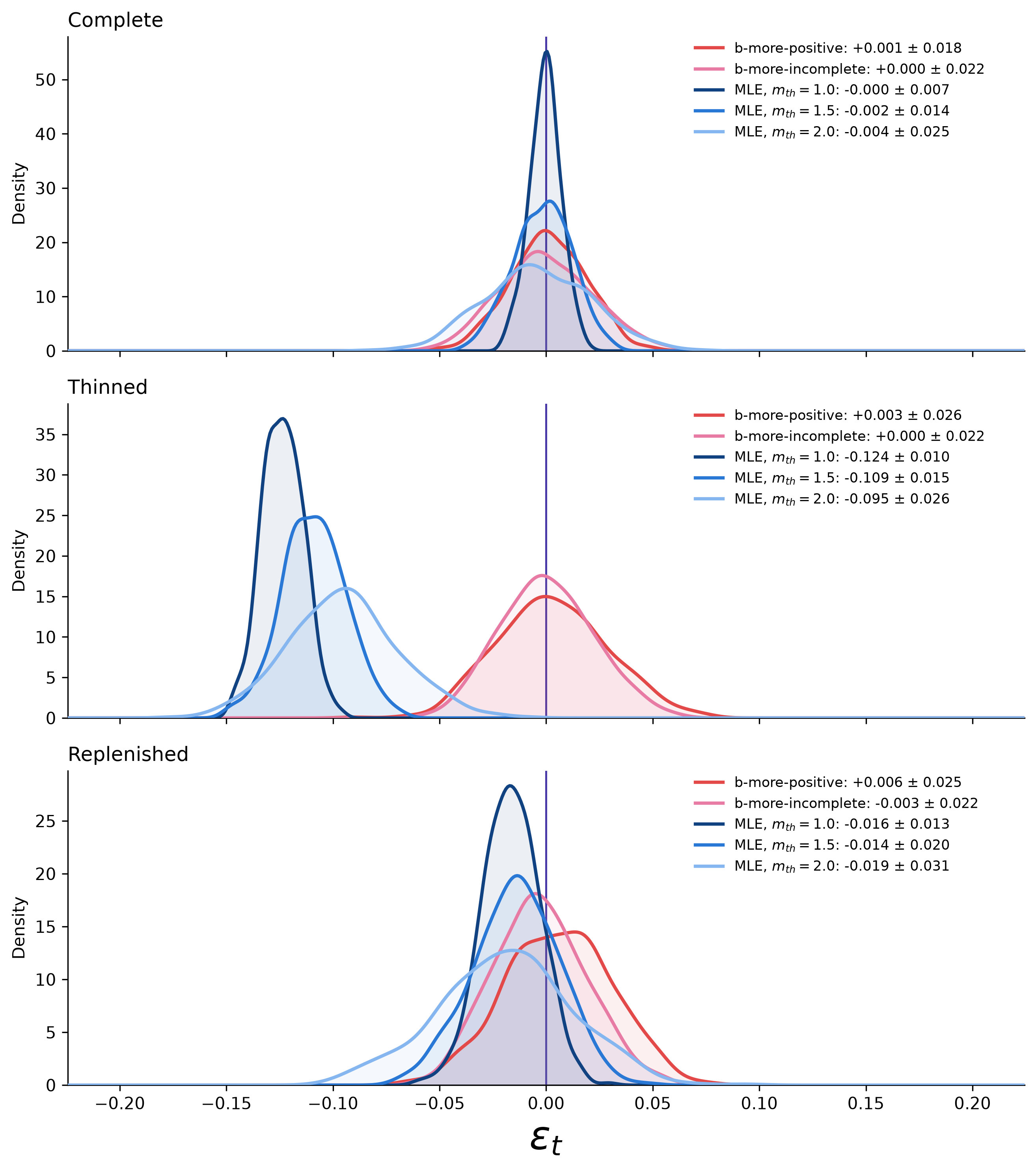}
\caption{Probability density of the relative $b$-value error
$\epsilon_t=\widehat{b}/b_{\mathrm{true}}-1$. Complete- and thinned-
catalog distributions are based on all 500 synthetic realizations,
whereas replenished-catalog distributions are based on the 484
realizations for which AutoBETA returned a valid depleted-region polygon. Results are shown for the complete catalogs (top), ETASI-thinned catalogs (middle), and AutoBETA-replenished catalogs (bottom). Red and pink curves represent the $b$-more-positive and $b$-more-incomplete estimators, respectively. Blue curves show the conventional maximum-likelihood estimates obtained using $M_{\mathrm{th}}=1.0$, 1.5, and 2.0. The vertical line marks the unbiased value $\epsilon_t=0$. Legend values report the mean and standard deviation across realizations.}
\label{fig:synthetic_b_errors}
\end{figure}

A value of $\epsilon_t=0$ therefore corresponds to exact recovery of the imposed Gutenberg--Richter slope. Figure~\ref{fig:synthetic_b_errors} presents the distributions of $\epsilon_t$ for the complete, thinned, and replenished catalogs.

All estimators are consistent with the theoretical value when applied to the complete catalogs. The mean relative errors are $0.002\pm0.025$ for the $b$-more-positive estimator and $-0.003\pm0.021$ for the $b$-more-incomplete estimator. The conventional maximum-likelihood estimator also performs as expected, with mean relative errors of $0.000\pm0.007$, $-0.002\pm0.014$, and $-0.004\pm0.025$ for $M_{\mathrm{th}}=1.0$, 1.5, and 2.0, respectively. The increasing dispersion at larger thresholds results from the progressively smaller number of earthquakes retained in the estimate.

The effect of incompleteness is markedly different among the estimators. In the thinned catalogs, the conventional estimator systematically underestimates the true $b$-value. The mean relative errors are $-0.124\pm0.010$, $-0.109\pm0.015$, and $-0.095\pm0.026$ for $M_{\mathrm{th}}=1.0$, 1.5, and 2.0, respectively. Raising the magnitude threshold reduces the magnitude of the bias, but it does not eliminate it within the investigated range and produces a larger statistical dispersion.

In contrast, the positive-difference estimators remain centered close to the theoretical value even before replenishment. The $b$-more-positive estimator gives $\epsilon_t=0.005\pm0.036$, while the $b$-more-incomplete estimator gives $\epsilon_t=-0.003\pm0.022$. The latter combines negligible bias with a comparatively narrow distribution, confirming its robustness to the ETASI detection losses.

Replenishment strongly improves the conventional maximum-likelihood estimates. After the missing population is reconstructed, the mean relative errors become $-0.016\pm0.013$, $-0.014\pm0.020$, and $-0.019\pm0.031$ for the three magnitude thresholds. The approximately 10--12\% bias observed in the thinned catalogs is therefore reduced to about 1--2\%. A small negative residual remains, consistent with the fact that the replenished catalogs contain fewer low-magnitude earthquakes than the corresponding complete realizations.

The positive-difference estimators remain stable after replenishment. Their mean relative errors are $0.007\pm0.033$ for $b$-more-positive and $-0.007\pm0.022$ for $b$-more-incomplete. Replenishment therefore does not substantially improve these estimates because they were already robust in the incomplete catalogs. At the same time, the procedure does not introduce a significant additional bias.

These results clarify the relationship between the two strategies. AutoBETA substantially restores the missing seismicity and makes the conventional Gutenberg--Richter estimate much more reliable. Positive-difference estimators instead recover the imposed slope directly from the incomplete catalog. The two approaches are therefore complementary: replenishment improves the statistical representation of the sequence as a whole, whereas the robust estimators protect a specific parameter of interest from detection-induced bias.

\section{Application to Southern California Aftershock Sequences}
\label{sec:real_data}

\begin{figure}[htbp]
\centering
\includegraphics[width=.78\textwidth]{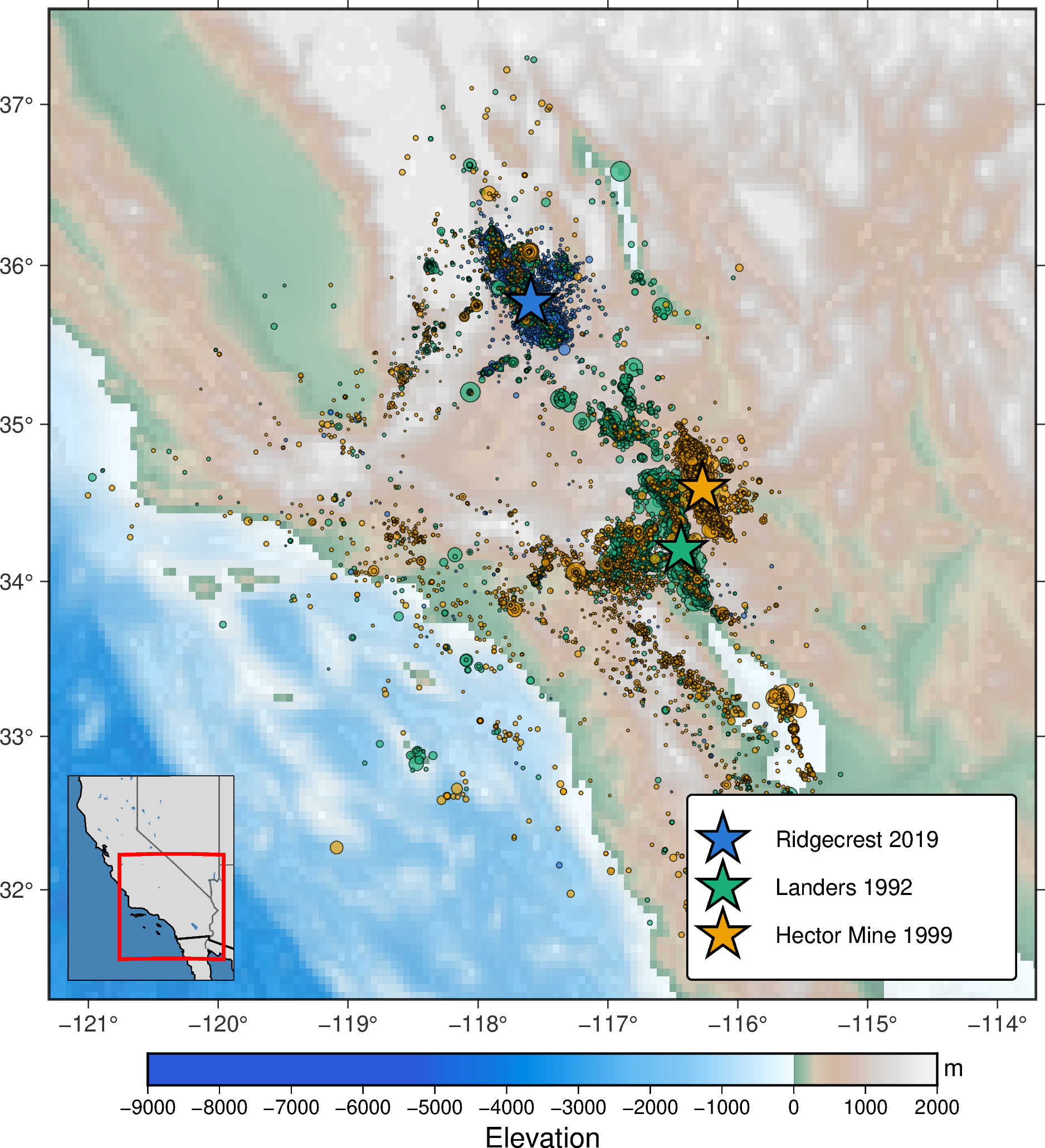}
\caption{Spatial distribution of the Southern California earthquake sequences analyzed in this study. Blue, green, and orange circles represent earthquakes associated with the 2019 Ridgecrest, 1992 Landers, and 1999 Hector Mine sequences, respectively, and symbol sizes scale with magnitude. Stars mark the corresponding mainshocks. Background colors indicate surface elevation and bathymetry in meters. The inset shows the location of the mapped region within California.}
\label{fig:socal_map}
\end{figure}

To investigate whether the conclusions obtained from the synthetic experiments also hold for real seismicity, we analyzed three well-known aftershock sequences from Southern California: the 1992 Landers ($M_w7.3$), the 1999 Hector Mine ($M_w7.1$), and the 2019 Ridgecrest ($M_w7.1$) earthquakes. These sequences were selected because they contain large numbers of aftershocks and exhibit pronounced short-term aftershock incompleteness immediately following their respective mainshocks. Figure~\ref{fig:socal_map} shows the geographical distribution of the three study regions. Although the sequences occurred under different tectonic and network conditions, all three present a characteristic depletion of small earthquakes during the earliest stages of the aftershock sequence.

For each sequence, the same replenishment procedure and the same three $b$-value estimators used in the synthetic analysis were applied without modification. This allows a direct comparison between the behavior observed under controlled ETAS simulations and that obtained from instrumental earthquake catalogs.

\subsection{Earthquake Replenishment}

\begin{figure}[htbp]
\centering
\includegraphics[width=0.94\textwidth]{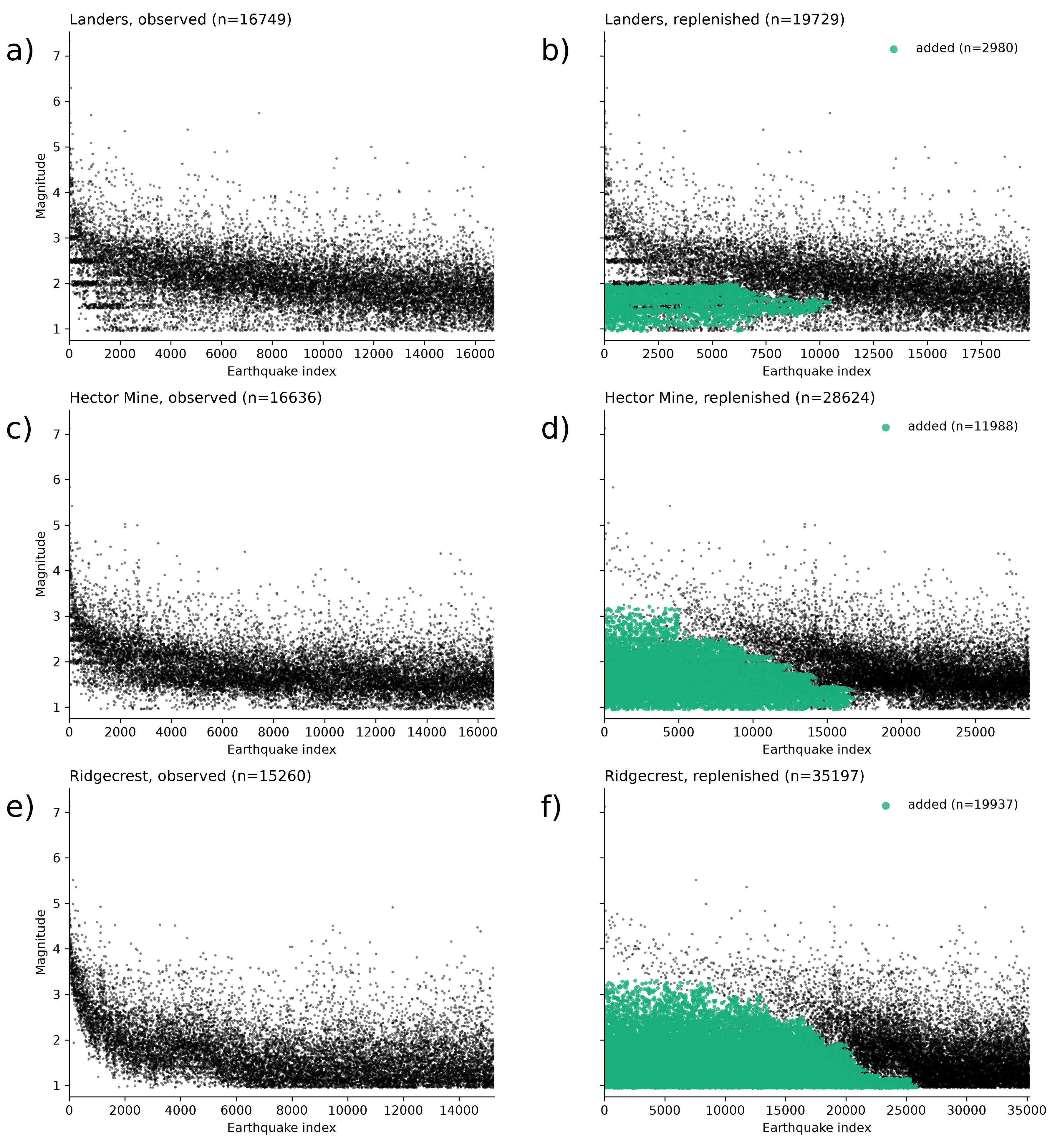}
\caption{Original and AutoBETA-replenished catalogs for the Landers, Hector Mine, and Ridgecrest aftershock sequences. Panels (a), (c), and (e) show the original observed catalogs for Landers, Hector Mine, and Ridgecrest, respectively. Panels (b), (d), and (f) show the corresponding replenished catalogs, with generated earthquakes highlighted in green. Landers increases from 16,749 to 19,719 earthquakes, Hector Mine from 16,636 to 28,439 earthquakes, and Ridgecrest from 14,308 to 32,421 earthquakes. The declining upper boundary of the reconstructed low-magnitude population reflects the
progressive recovery of earthquake detectability during each sequence.}
\label{fig:real_catalogs}
\end{figure}

Figure~\ref{fig:real_catalogs} compares the observed and replenished catalogs for the three aftershock sequences. In all cases, AutoBETA reconstructs additional earthquakes primarily during the early stages of the sequence and at low magnitudes, consistent with the expected signature of short-term aftershock incompleteness.

The amount of replenishment differs substantially among the three sequences. Ridgecrest exhibits the strongest correction, with 18,113 additional earthquakes generated from an original catalog containing 14,308 events. Hector Mine is similarly affected, increasing from 16,636 to 28,439 earthquakes after replenishment. By contrast, the Landers sequence requires a considerably smaller correction, increasing from 16,749 to 19,719 earthquakes. Despite these differences in the total number of reconstructed events, the generated earthquakes occupy the same region of the magnitude-time space in all three sequences, namely the low-magnitude portion immediately following the mainshock. This behavior agrees closely with that observed in the synthetic experiments and supports the interpretation that the replenishment algorithm primarily compensates for short-term detection losses. 

\subsection{Comparison of the $b$-Value Estimates}

The resulting $b$-value estimates are summarized in Table~\ref{tab:real_bvalues}. As expected, the conventional maximum-likelihood estimator systematically increases after replenishment for all three sequences, reflecting the recovery of a substantial number of previously missing small earthquakes.

\begin{table}[htbp]
\caption{$b$-Value Estimates for the Original and Replenished Southern
California Aftershock Sequences}
\label{tab:real_bvalues}
\centering
\resizebox{\textwidth}{!}{
\begin{tabular}{l cc cc cc}
\hline
& \multicolumn{2}{c}{Ridgecrest}
& \multicolumn{2}{c}{Landers}
& \multicolumn{2}{c}{Hector Mine} \\
\cline{2-3}
\cline{4-5}
\cline{6-7}
Estimator
& Original & Replenished
& Original & Replenished
& Original & Replenished \\
\hline
$b$-more-positive
& $1.004 \pm 0.011$
& $0.857 \pm 0.006$
& $1.032 \pm 0.010$
& $0.899 \pm 0.008$
& $1.158 \pm 0.012$
& $1.028 \pm 0.008$ \\
$b$-more-incomplete
& $0.970 \pm 0.015$
& $0.971 \pm 0.015$
& $1.044 \pm 0.011$
& $1.018 \pm 0.010$
& $1.141 \pm 0.010$
& $1.107 \pm 0.010$ \\
MLE ($M_{\mathrm{th}}=1.0$)
& $0.548 \pm 0.004$
& $0.660 \pm 0.003$
& $0.392 \pm 0.002$
& $0.422 \pm 0.002$
& $0.503 \pm 0.002$
& $0.564 \pm 0.002$ \\
MLE ($M_{\mathrm{th}}=1.5$)
& $0.679 \pm 0.006$
& $0.815 \pm 0.006$
& $0.593 \pm 0.004$
& $0.644 \pm 0.004$
& $0.797 \pm 0.006$
& $0.914 \pm 0.006$ \\
MLE ($M_{\mathrm{th}}=2.0$)
& $0.737 \pm 0.010$
& $0.881 \pm 0.010$
& $0.861 \pm 0.008$
& $0.861 \pm 0.008$
& $0.955 \pm 0.013$
& $1.065 \pm 0.013$ \\
\hline
\end{tabular}
}
\begin{tablenotes}
\item Reported uncertainties are bootstrap standard errors from 100
replicates conditional on the displayed catalog realization. Variability
among independent replenishment draws is reported separately in
Figure~\ref{fig:seed_sensitivity}.
\end{tablenotes}
\end{table}

Replenishment increases the conventional maximum-likelihood estimate in
nearly all sequence--threshold combinations; the exception is Landers at
$M_{\mathrm{th}}=2.0$, for which the reported estimate is unchanged.
The correction is strongest for Ridgecrest and remains strongly dependent
on the imposed magnitude threshold.

The two positive-difference estimators respond differently to the
generated events. The $b$-more-positive estimate decreases by approximately
11--12\% after replenishment in all three sequences: from 1.020 to 0.894
for Ridgecrest, from 1.058 to 0.929 for Landers, and from 1.203 to 1.070
for Hector Mine. It should therefore not be described as invariant to
replenishment in the instrumental catalogs. This behavior likely reflects
the sensitivity of its selected earthquake pairs to the insertion and
chronological ordering of generated events.

The $b$-more-incomplete estimator is substantially more stable. Its
changes are approximately $-0.3\%$, $-3.0\%$, and $-3.1\%$ for Ridgecrest,
Landers, and Hector Mine, respectively. Because the true slopes of the
instrumental sequences are unknown, these comparisons evaluate
methodological stability rather than estimator bias. The repeated-seed
analysis in Figure~\ref{fig:seed_sensitivity} further shows that the
differences among the estimators are larger than the variability caused
by the stochastic replenishment draw.

\subsection{Stochastic Replenishment Stability}
\label{subsec:stochastic_stability}

To quantify the uncertainty introduced by the stochastic generation of
missing events, AutoBETA was repeated $S=10$ times for each instrumental
catalog using independent random seeds. The observed input catalog and all
algorithmic parameters were held fixed across repetitions. For every draw
$s$, we recorded the number of generated events $N_{\mathrm{add}}^{(s)}$
and recomputed the five $b$-value estimates: $b$-more-positive,
$b$-more-incomplete, and the maximum-likelihood estimates for
$M_{\mathrm{th}}=1.0$, 1.5, and 2.0.
The estimator settings were also kept fixed, including
$\delta M_{\mathrm{th}}=0.2$ and $R_{\max}=60$~km.

The stochastic-stability experiment is shown in
Figure~\ref{fig:seed_sensitivity}. The number of generated events varies
among replenishment draws, particularly for Ridgecrest and Hector Mine.
Nevertheless, the within-sequence distributions of the five $b$-value
estimates are narrow relative to the differences among estimators. The
systematic decrease of the $b$-more-positive estimate after replenishment
is therefore not explained solely by an anomalous random seed. The
$b$-more-incomplete estimator remains the closest to its estimate from
the original catalog, whereas the conventional estimates retain a strong
dependence on the adopted magnitude threshold.

\section{Discussion and Conclusions}
\label{sec:discussion_conclusions}

This study compared two conceptually different strategies for addressing short-term aftershock incompleteness. Earthquake replenishment seeks to reconstruct a statistically plausible population of undetected events, thereby producing a catalog that can support a broad range of seismological analyses. Positive-difference estimators instead target a specific statistical quantity,
the Gutenberg-Richter $b$-value, without attempting to reconstruct the missing earthquakes. Our results show that these approaches are closely related but
should not be regarded as statistically equivalent. The synthetic experiments provide the clearest assessment because both the complete catalog and the imposed Gutenberg-Richter slope are known. The ETASI detection model removes a large fraction of small earthquakes and produces the
expected time-dependent deficit of low-magnitude events. Under these conditions, the conventional maximum-likelihood estimator underestimates the true $b$-value by approximately 10-12\%, depending on the adopted magnitude
threshold. Increasing $M_{\mathrm{th}}$ reduces the bias only partially and simultaneously increases the statistical dispersion because fewer earthquakes remain available for estimation. This illustrates a fundamental limitation of
threshold-based analyses during rapidly evolving aftershock sequences: the threshold must be sufficiently high to avoid incompleteness, but increasing it also reduces the information available for resolving the magnitude distribution. \\
AutoBETA reconstructs the broad geometry of the missing-event population and substantially improves the conventional maximum-likelihood estimates. After replenishment, the mean residual error is reduced to approximately 1-2\% for all three magnitude thresholds considered. This is an important result because
it demonstrates that replenishment can recover enough of the low-magnitude population to restore most of the information lost through short-term incompleteness. The method therefore represents a valuable approach when the
scientific objective depends on the occurrence times, number, or magnitude distribution of the missing earthquakes. \\
The small residual bias observed after replenishment should not be interpreted as evidence that the replenishment procedure fails. AutoBETA does not attempt to reproduce the specific earthquakes removed from an individual catalog
realization. Rather, it infers the depleted region from the observed point pattern and generates a population that is statistically compatible with that region. Consequently, the number and magnitudes of replenished earthquakes need
not coincide exactly with the original missing population. In our synthetic example, the reconstructed catalog remains smaller than the corresponding complete catalog, with the remaining deficit concentrated primarily among low-magnitude earthquakes. Because the Gutenberg-Richter slope is particularly sensitive to the relative number of events close to the lower magnitude threshold, even a comparatively small mismatch in this part of the distribution can produce a measurable residual bias.\\
The behavior of the positive-difference estimators is fundamentally different. Both the $b$-more-positive and $b$-more-incomplete estimators remain centered close to the imposed $b$-value in the complete, thinned, and replenished
catalogs. Their stability before replenishment is consistent with their design: they seek to reduce the effect of short-term incompleteness through conditional
magnitude differences rather than by reconstructing the missing population. Replenishment therefore provides little additional improvement when these estimators are already applied under conditions for which they are robust. At
the same time, their estimates remain stable after new earthquakes are added, showing that the replenishment procedure does not generate a substantial additional bias in the synthetic experiment.\\
Among the two positive-difference approaches, the $b$-more-incomplete estimator shows the smallest sensitivity to replenishment. This behavior is expected from
the artificial blind-time filter, which removes the part of the observed catalog most likely to retain a dependence on previous larger earthquakes. The $b$-more-positive estimator also recovers the true value in the synthetic
catalogs, although its distribution is broader and its estimates vary more after replenishment in the instrumental applications. These differences indicate that the two estimators should not be used interchangeably without
considering their respective parameter choices and assumptions.\\
The Southern California applications provide an observational counterpart to
the controlled experiments. AutoBETA reconstructs substantial populations of low-magnitude earthquakes in the Ridgecrest and Hector Mine sequences and a
smaller population for Landers. In all three cases, the conventional maximum-likelihood estimates increase after replenishment, as expected when previously missing small earthquakes are restored. The size of the correction
varies strongly among sequences and magnitude thresholds, reflecting differences in apparent incompleteness as well as in the geometry and temporal evolution of the reconstructed regions. \\
The instrumental results must nevertheless be interpreted more cautiously than the synthetic results because the true $b$-values and the actual missing-event populations are unknown. An increase after replenishment does not by itself
prove that the resulting estimate is unbiased. Similarly, the agreement of two estimators does not establish that either one represents the exact underlying magnitude distribution. The real-catalog analysis instead provides a test of methodological stability. In this respect, the $b$-more-incomplete estimates change only slightly after replenishment, whereas the conventional estimates and, to a lesser extent, the $b$-more-positive estimates are more sensitive to the generated population. This behavior is consistent with the controlled synthetic results and supports the use of the blind-time estimator when the
principal objective is a robust estimate of the Gutenberg-Richter slope.\\
The distinction between the two strategies has practical implications. Replenishment is particularly useful when the missing earthquakes themselves are scientifically relevant. Examples include analyses of aftershock rates, Omori-law decay, triggering productivity, clustering, and point-process models whose likelihood depends on the number and occurrence times of events. For these applications, a robust $b$-value estimate alone cannot replace the
information supplied by a reconstructed catalog. Conversely, when the primary target is only the slope of the magnitude-frequency distribution, direct use
of a robust positive-difference estimator may avoid the additional uncertainty associated with generating a missing-event population. The preferred approach should therefore be determined by the inferential objective rather than by a
general ranking of the methods.\\
Several limitations should be considered. First, the synthetic assessment is based on one form of rate and magnitude-dependent incompleteness generated
through the ETASI model. Although this formulation reproduces the principal features of short-term aftershock incompleteness, other detection mechanisms, network geometries, and magnitude-error models could affect the relative performance of the methods. Second, both replenishment and the positive-difference estimators rely, either explicitly or effectively, on the assumption that the underlying magnitude distribution is separable from earthquake occurrence times. Genuine temporal variations of the $b$-value could therefore be difficult to distinguish from detection effects and should be investigated in future controlled experiments. Third, the instrumental analysis is limited to three Southern California sequences. Additional applications to other tectonic settings, networks, and catalog types will be necessary to evaluate the generality of the observed behavior. A further limitation concerns the geometry required by the replenishment
procedure. Even for the deterministic thinning model, AutoBETA failed to
identify a valid BEPIT missing-region polygon in 16 of 500 realizations
(3.2\%). Moreover, preliminary experiments with a gradual detection
transition, $\sigma>0$, produced diffuse depletion patterns for which a
valid connected polygon could not be recovered reliably. The reported
replenishment performance is therefore conditional both on successful
polygon construction and on the hard-threshold incompleteness model.
Extending the method to smoothly varying detection probabilities remains
an important direction for future work.

In conclusion, earthquake replenishment markedly improves the statistical representation of incomplete aftershock sequences and substantially reduces the bias of conventional maximum-likelihood $b$-value estimates. However, recovering a more complete catalog and obtaining an unbiased estimate of the Gutenberg-Richter slope are related but distinct statistical objectives. In the synthetic experiments, replenishment removes most, but not all, of the bias affecting conventional estimates, whereas the $b$-more-positive and $b$-more-incomplete estimators recover the imposed slope directly from the
incomplete catalogs. The instrumental applications show the same qualitative contrast, with the $b$-more-incomplete estimator displaying the greatest stability before and after replenishment. Earthquake replenishment and robust
$b$-value estimation should therefore be viewed as complementary tools: replenishment reconstructs the missing seismicity for general-purpose catalog analysis, while robust estimators provide a direct solution when the primary
objective is inference on the Gutenberg-Richter $b$-value.

\appendix

\section{Stochastic Stability Across Replenishment Realizations}
\label{app:stochastic_stability}

\begin{figure}[htbp]
\centering
\includegraphics[width=\textwidth]{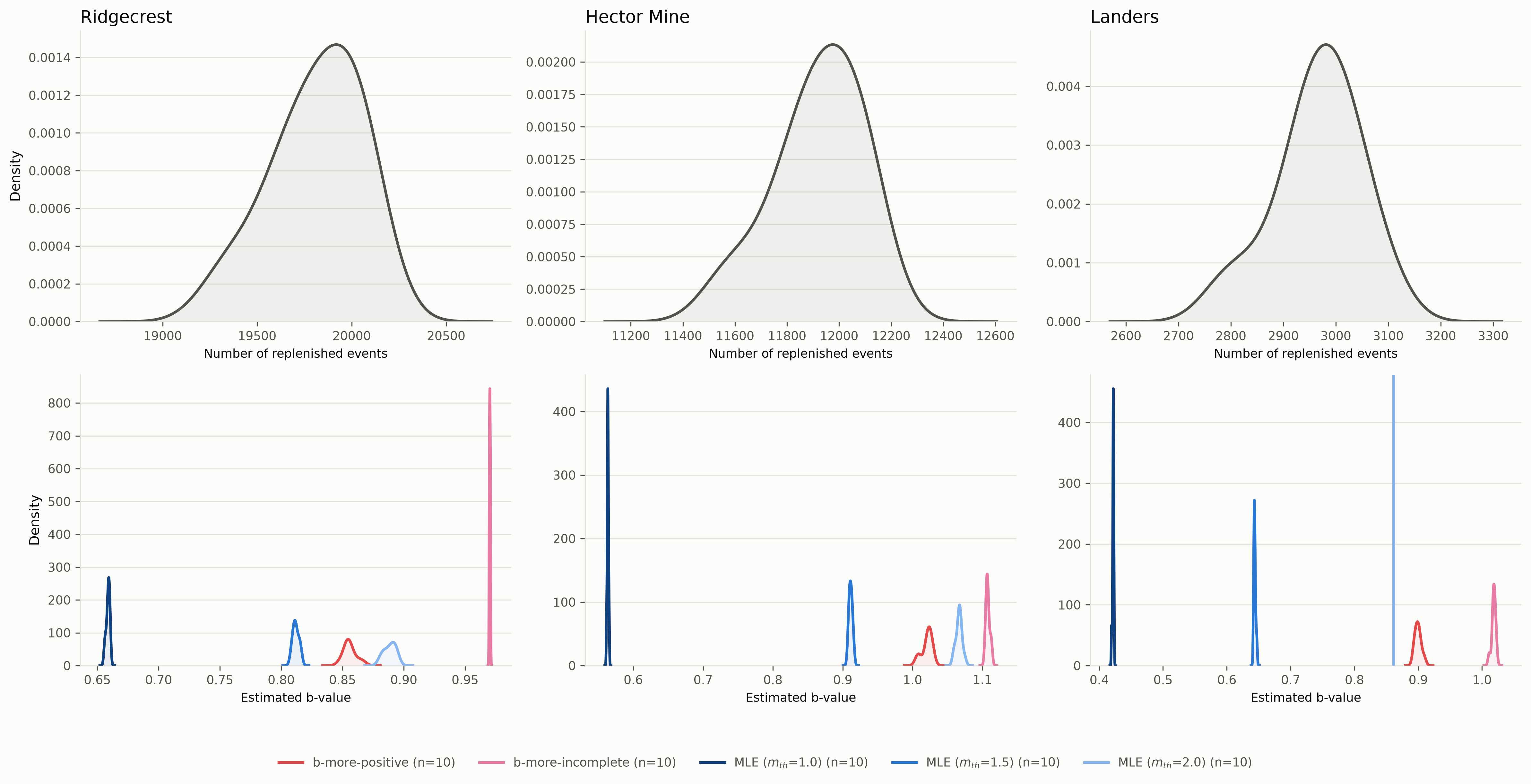}
\caption{Stochastic stability of the instrumental-catalog results across
10 independent replenishment seeds. Top panels show kernel-density
summaries of the number of synthetic events added to the Ridgecrest,
Hector Mine, and Landers catalogs. Bottom panels show the corresponding
distributions of the $b$-more-positive, $b$-more-incomplete, and
maximum-likelihood estimates at $M_{\mathrm{th}}=1.0$, 1.5, and 2.0.
The observed catalog, AutoBETA settings, and estimator parameters,
including $\delta M_{\mathrm{th}}=0.2$ and $R_{\max}=60$~km, were held
fixed. The figure therefore represents sensitivity to the stochastic
replenishment draw, not a hyperparameter-sensitivity analysis.}
\label{fig:seed_sensitivity}
\end{figure}

Figure~\ref{fig:seed_sensitivity} summarizes 10 independent AutoBETA
replenishment draws for each instrumental sequence. These repetitions
quantify the Monte Carlo variability associated with generating a
statistically plausible missing-event population while keeping the
observed catalog and every tuning parameter fixed. They do not test
sensitivity to alternative grid resolutions, deficient-cell percentiles,
connected-region thresholds, or estimator parameters.

\section*{Open Research Section}
{\sloppy The code for the Auto‐Biscale Empirical Transformation Application (AutoBETA) and example datasets are available at \url{https://github.com/zhancx-edu/EarthquakeCatalogReplenishment}. The b-more-positive code is freely available at \url{https://doi.org/10.5281/zenodo.19198292}.\par}

\section*{Conflict of Interest declaration}
The authors declare there are no conflicts of interest for this manuscript.

\acknowledgments
G.P. acknowledges support from the DyDéCo project from the CPJ Chair at ENS de Lyon and CNRS.

\bibliography{Reference.bib}

\end{document}